\documentclass[smallextended]{svjour3}                 
\smartqed 
\usepackage{graphicx}
\newcommand{\url}[1]{[{\tt #1}]}
\newcommand{\dd}{{\rm d}}
\newcommand{\ppn}{{\rm PPN}}
\newcommand{\dark}{{\rm de}}
\newcommand{\de}{{\rm de}}
\newcommand{\mat}{{\rm m}}
\newcommand{\dyn}{{\rm dyn}}
\newcommand{\lens}{{\rm lens}}
\newcommand{\etal}{{\em et al.}}
\newcommand{\br}{{\bf r}}
\newcommand{\bk}{{\bf k}}
\newcommand{\bt}{{\bf \theta}}

\begin{document}

\title{Tests of General Relativity on Astrophysical Scales }

\author{Jean-Philippe Uzan}

\institute{Jean-Philippe Uzan \at
              Institut d'Astrophysique de Paris, 
              UMR-7095 du CNRS,
              Universit\'e Pierre et Marie Curie, Paris VI,  \\
              98bis bd. Arago, 75014 Paris, France.\\
              Tel.: +33-1-44328026\\
              Fax: +33-1-44328001\\
              \email{uzan@iap.fr}}

\date{Received: date / Accepted: date}
\maketitle
\begin{abstract}
While tested to a high level of accuracy in the Solar system, general
relativity is under the spotlight of both theoreticians and
observers on larger scales, mainly because of the need to
introduce dark matter and dark energy in the cosmological model.
This text reviews the main tests of general relativity focusing
on the large scale structure and more particularly weak lensing.
The complementarity with other tests (including those on Solar
system scales and the equivalence principle) is discussed.

\keywords{General Relativity\and Cosmology \and Weak lensing}
\end{abstract}

\section{Introduction}\label{intro}

Gravitational lensing is historically bound to the developments of general
relativity (GR) and, more generally, of the theories of gravitation. Since the end
of the 18th century, it was thought that light can be deflected by a gravitational
field, in particular with the works of Georg von Soldner that postulated that light
must behave as any other particle or of Robert Blair, John Mitchell
and Pierre Simon de Laplace (see Ref.~\cite{eisen} for an historical discussion).

The deflection of light by any massive body is a central prediction of GR.
In particular, the observations of the deflection of light emitted by
distant stars by the Sun during the Solar eclipse on the 29th
May 1919 by the expeditions led by Eddington and Cottingham on Principe island
and by Davidson and Crommelin in the Nordeste region of Brasil is always considered
as an experimental confirmation of the predictions of GR. Indeed, if such
an observation was a test of GR, it is only because the mass
of the Sun was supposed to be well-determined at the time. On the one hand, the 
light deflection angle predicted by GR is
$$
  \Delta\theta_{\rm GR} = \frac{4 GM_\odot}{bc^2}
$$
where $G$ is the Newton constant, $M_\odot$ the Solar mass,
$b$ the impact parameter and $c$ the speed of light, while, 
on the other hand, the
dynamics of massive bodies, such as planets, are in an extremely
good approximation still given by Kepler third law,
$$
  \frac{P}{2\pi} = \frac{a^3}{GM_\odot},
$$
where $P$ is the period of the orbit and $a$ its semi-major axis. Measuring
$\Delta\theta$ and $b$ on one side and $P$ and $a$ on the other gives
two estimates of $GM_\odot$,
$$
  GM^\lens_\odot = \frac{bc^2\Delta\theta}{4}, \qquad
   GM^\dyn_\odot = \frac{2\pi a^3}{P}, 
$$
that needs, given the error bars, to be consistent. This illustrates that lensing {\em alone} does not
allow to construct a test of the theory of gravity but that we need to
check the consistency between various predictions such as
lensing and dynamics of massive bodies.

Today, gravity, i.e. the only long range force that cannot be screened, is described
by GR in which it is the consequence of the geometry of the spacetime.
General relativity is consistent with all precision experimental tests available but most
of these tests are restricted to the Solar system or to our Galaxy, so that they
are only local. By considering astrophysical systems, we can extend the {\em domain
of validity} of GR at large distance, low acceleration or low curvature.
In particular, most atttempts to construct a quantum theory of gravity or
to unify it with other interactions predict the existence of partners to the graviton, i.e.
extra fields contributing to a long range force, and thus to gravity (e.g. this
this the case in all extra-dimensional theories where some components
of the extra-dimensional part of the metric behave as scalar fields from
a 4-dimensional point of view; string theory also
predicts the existence of a scalar field, the dilaton, in the graviton supermultiplet).
It follows that deviation from GR~\cite{dn1} are expected but in
many cases (such as scalar-tensor theories), the theory
can be dynamically attracted toward GR so that cosmology can set
sharper constraints than those obtained locally.

An other reason to test GR in these regimes is related to our
current cosmological model. This model, which is consistent with almost
all astrophysical data requires the addition of both dark matter (a fluid
with negligible pressure that does not interact with standard matter) and dark
energy (a fluid with a negative pressure), which represent, respectively 23\% and 72\% of the matter 
content of the universe. The need for these two components arises from 
the study of the dynamics of
clusters, galaxies, large scale structures and of the cosmic expansion 
under the assumption
that GR holds on astrophysical scales. This conclusion has been
challenged by invoking possible modifications of GR either in
a low acceleration regime to explain the galaxy rotation curves and cluster
dynamics without introducing dark matter and at large distance to account
for the late time acceleration of the cosmic expansion.

In conclusion, the possibility to sharpen our understanding of the validity of GR
from astrophysical data and the need to understand the properties
of dark energy and dark matter, which are tied to the validity of GR,
are our two driving motivations. The main difficulty is that, on 
astrophysical scales, most observations entangle the properties
of gravity, matter, as well as other hypothesis such as the Copernican
principle.

The bottom line of the construction of these tests is simple. Once
GR is assumed valid, it describes the dynamics
of the cosmic expansion, the growth of large scale structures,
the propagation of light etc... so that many observables are
not independent. Such observables can be used to construct
consistency relations. Any sign of a violation of these
relations will indicate the need to extend our description
of gravity, but will not indicate us how. For instance, in the
oversimplified Solar system example described
above, we want the two estimates of $GM_\odot$ to agree that is
we must have
\begin{equation}\label{relSS}
  \frac{bc^2\Delta\theta}{4} - \frac{2\pi a^3}{P}=0.
\end{equation}
This is a relation between observable quantities ($b, \Delta\theta, P, a$) that has
to be satisfied. Actually, it was first proposed in Ref.~\cite{ub01} to perform a
similar test on cosmological scales using weak lensing and galaxy redshift suveys,
followed by the analysis of Ref.~\cite{wk01}.

The review is organized as follows. We start, in \S~\ref{sec:2}, by recalling the main 
hypothesis on which GR is based as well as the standard 
constraints obtained in the Solar system. We also discuss the use of alternative theories
and draw the conclusions of what was learnt in the Solar system for constructing tests on
astrophysical scales. In \S~\ref{sec:3}, we discuss briefly tests on galactic and
cluster scales where the need for dark matter can be interpreted as the
necessity to modify GR in a low acceleration regime. Larger scales are
considered in \S~\ref{sec:4} which focuses on the large scale structure
of the universe.

\section{Relativity and its Solar system tests}\label{sec:2}

\subsection{General relativity (in brief)}

Let us recall that GR, Einstein's theory of gravity, 
relies on two independent hypothesis.

First, the theory rests on the {\it Einstein equivalence principle},
which includes the universality of free fall, the local position
and local Lorentz invariances in its weak form (as other metric
theories) and is conjectured to satisfy it in its strong form. We
refer to Ref.~\cite{willbook} for a detailed description of these
principles and their implications. The weak equivalence principle
can be mathematically implemented by assuming that all matter
fields, including gauge bosons, are minimally coupled to a single metric tensor
$g_{\mu\nu}$. This metric defines the lengths and times measured by
laboratory rods and clocks so that it can be called the {\it
physical metric}. This implies that the action for any matter
field, $\psi$ say, is of the form 
\begin{equation}\label{graction1}
S_\mat[\psi,g_{\mu\nu}].
\end{equation}
This so-called {\it metric coupling}
ensures in particular the validity of the universality of
free-fall.

Then, the action for the gravitational sector is given by the
Einstein-Hilbert action
\begin{equation}\label{graction2}
 S_{\rm gravity} = \frac{c^3}{16\pi G}\int \dd^4x\sqrt{-g_*}R_*,
\end{equation}
where $g^*_{\mu\nu}$ is a massless spin-2 field called the {\it
Einstein metric}. The second hypothesis of GR states that both metrics
coincide, i.e.
$$
 g_{\mu\nu} = g^*_{\mu\nu}.
$$

\subsection{Testing GR}

It follows that one can aim at testing both the equivalence principle and
the dynamical equations that derive from the Einstein-Hilbert action.

The assumption of metric coupling is well tested in the Solar
system. First it implies that all non-gravitational constants are
spacetime independent, which have been tested to a very high
accuracy in many physical systems and for various fundamental
constants~\cite{uzan03,uzan04,uzan09,ctebook}, e.g.
at the $10^{-7}$ level for the fine structure constant on time
scales ranging to 2-4 Gyrs. Second, the isotropy has been tested
from the constraint on the possible quadrupolar shift of nuclear
energy levels~\cite{isotest,isotest2,isotest3} proving that matter couples to a unique metric tensor
at the $10^{-27}$ level. Third, the universality of free fall of
test bodies in an external gravitational field at the $10^{-13}$
level in the laboratory~\cite{uff1,uff2}. The Lunar Laser ranging experiment~\cite{uff3},
which compares the relative acceleration of the Earth and Moon in
the gravitational field of the Sun, also probe the strong
equivalence principle at the $10^{-4}$ level. Fourth, the Einstein
effect (or gravitational redshift) states that two identical
clocks located at two different positions in a static Newton
potential $U$ and compared by means of electromagnetic signals
shall exhibit a difference in clock rates of $1+[U_1-U_2]/c^2$.
This effect has been
measured at the $2\times10^{-4}$ level~\cite{clock1}.

The parameterized post-Newtonian formalism (PPN) is a general
formalism that introduces 10 phenomenological parameters to
describe any possible deviation from GR at the
first post-Newtonian order~\cite{willbook}. The formalism assumes
that gravity is described by a metric and that it does not involve
any characteristic scale. In its simplest form, it reduces to the
two Eddington parameters entering the metric of the Schwartzschild
metric in isotropic coordinates
$$
 g_{00} = - 1 + \frac{2Gm}{rc^2} -
 2\beta^\ppn\left(\frac{2Gm}{rc^2}\right)^2,
 \qquad
 g_{ij} = \left(1+2\gamma^\ppn\frac{2Gm}{rc^2}\right)\delta_{ij}.
$$
Indeed, general relativity predicts $\beta^\ppn=\gamma^\ppn=1$.
These two phenomenological parameters are constrained (1) by the
shift of the Mercury perihelion~\cite{mercure} which
implies that $|2\gamma^\ppn-\beta^\ppn-1|<3\times10^{-3}$, (2) the
Lunar laser ranging experiments~\cite{uff3} which
implies $|4\beta^\ppn-\gamma^\ppn-3|=(4.4\pm4.5)\times10^{-4}$ and
(3) by the deflection of electromagnetic signals which are all
controlled by $\gamma^\ppn$. For instance the very long baseline
interferometry~\cite{vlbi} implies that
$|\gamma^\ppn-1|=4\times10^{-4}$ while the measurement of the time
delay variation to the Cassini spacecraft~\cite{cassini}
sets $\gamma^\ppn-1=(2.1\pm2.3)\times10^{-5}$.

The PPN formalism does not allow to test finite range effects that
could be caused e.g. by a massive degree of freedom. In that case
one expects a Yukawa-type deviation from the Newton potential,
$$
 \Phi=-\frac{Gm}{r}\left[1+\alpha\left(1-\hbox{e}^{-r/\lambda}\right)\right],
$$
that can be probed by ``fifth force'' experimental searches.
$\lambda$ characterizes the range of the Yukawa deviation while
its strength $\alpha$ may also include a
composition-dependence~\cite{uzan03}. The constraints on
$(\lambda,\alpha)$ are summarized in Ref.~\cite{5force} which
typically shows that $\alpha<10^{-2}$ on scales ranging from the
millimeter to the Solar system size.

GR is also tested  with pulsars~\cite{gefpul2,gefpul} and in the
strong field regime~\cite{psaltis}. For more details we refer to
Refs.~\cite{willbook,lilley,tury}.
Needless to say that any extension of GR has to
pass these constraints. However, these deviations
can be larger in the past, as we shall see, which makes
cosmology an interesting field to extend these
constraints.

\subsection{Alternative theories of gravity}

The ways of modifying GR are so
various and in large number that we cannot review
them here. We refer to Refs.~\cite{willbook,gefbruneton}
for some examples. 

Let us however introduce the simplest modification of GR, that
is scalar-tensor theories of gravity in which gravity is mediated not only by a
massless spin-2 graviton but also by a spin-0 scalar field that
couples universally to matter fields (this ensures the
universality of free fall). Their action can be
written as, in the so-called Einstein frame,
\begin{eqnarray}\label{STaction}
 S &=& \frac{1}{16\pi G_*}\int \dd^4x\sqrt{-g_*}\left[ R_*
        -2g_*^{\mu\nu} \partial_\mu\varphi_*\partial_\nu\varphi_*
        - 4V(\varphi_*)\right]\nonumber\\
   && \qquad + S_{\rm matter}[A^2(\varphi_*)g^*_{\mu\nu};\psi],
\end{eqnarray}
where $G_*$  is the bare gravitational constant. The physical
metric, to which matter is universally coupled, 
$g_{\mu\nu}=A^2(\varphi_*)g^*_{\mu\nu}$ is
the product of the coupling function $A$, which characterizes the
strength of the scalar interaction, and the Einstein frame metric
$g^*_{\mu\nu}$. This theory involves a new degree of freedom
coupled to matter.

It can be used to illustrate the effect of modification of GR on lensing
(see Ref.~\cite{gefbruneton} for more details).
Consider the action for electromagnetism in $d$ dimensions
$$
 S_{\rm Maxwell} = \frac{1}{4}\int \sqrt{-g_*}
 g_*^{\mu\nu}g_*^{\rho\sigma}F_{\mu\rho}F_{\nu\sigma}\dd^d x
$$
transforms to
$$
 S_{\rm Maxwell} = \frac{1}{4}\int \sqrt{-g}
 g^{\mu\nu}g^{\rho\sigma}A^{d-4}F_{\mu\rho}F_{\nu\sigma}\dd^d x
$$
under the confomal transformation $g_{\mu\nu}=A^2(\varphi_*)g^*_{\mu\nu}$.
In the relevant case of a $d=4$ dimensional spacetime, the Maxwell
action is conformally invariant. Therefore light is only coupled to the
spin-2 field $g^*_{\mu\nu}$ so that light deflection by a point mass
$M$ must be the same as in GR, i.e.
$$
  \Delta\theta = \frac{4 G_*MA^2}{bc^2},
$$
where $A^2M$ is the deflecting mass in the Einstein frame.
It thus seems that there is no effect on lensing, contrary to the standard lore
that light deflection is smaller in scalar-tensor theories. Actually there is
a crucial difference since in scalar-tensor theory massive
bodies do feel the scalar field. It follows that the gravitional constant measured
in a Cavendish experiment today is not $G_*$ but $G_N = G_*A_0^2(1+\alpha_0^2)$
with $\alpha\equiv \dd\ln A/\dd\varphi_*$ and where a subscript 0
indicates that the quantity is evaluated today. It follows that the dynamics of massive
bodies, such as planetary orbits, determine $G_NM$ and not $G_*MA^2$
so that 
$$
  \Delta\theta = \frac{4 G_*MA_0^2}{bc^2}=
  \frac{4 G_NM}{(1+\alpha_0^2)bc^2} = \frac{\Delta\theta_{\rm GR}}{1+\alpha_0^2} \leq
  \Delta\theta_{\rm GR},
$$
as expected. Again, this shows that lensing {\it alone} cannot probe GR
and that we need to compare different measurements. 
Note also that the gravitational constant (or more
precisely the dimensionless number $Gm^2/\hbar c$) varies with time
so that extending this argument to the cosmological context is not straightforward~\cite{sur04}.
When the theoretical framework is specified then the post-Newtonian parameters
can be computed (here $\gamma^\ppn=-2\alpha^2/(1+\alpha^2)$
and $\beta^\ppn=\alpha^2/[2(1+\alpha^2)^2]\dd\alpha/\dd\varphi_*$ as
long as the potential is such that the field is light
on Solar system scales) so that
the PPN constraints can be translated to constraints on the parameters
of the model.

In conclusion, this simple extension of GR illustrates that 
we always have to introduce new fields in the theory so that we
have to specify their nature (here a scalar field) and the ways they
couple to the matter fields (here universally with the strength
$\alpha$). The distinction between a modification of GR and
dark matter (or energy) is thus slight since in {\it both cases we need
to introduce new fields in our theory}. The main difference lies
in the fact that the amount of dark matter or dark energy is
set by initial conditions (e.g. the amount of dark matter is 
fixed initially and determines the properties of the potential wells
in which baryonic matter falls to form the structure or
the amount of dark energy is fixed by tuning some parameters
and/or initial conditions so that it starts dominating today and the
fact that $\rho_\Lambda:\rho_{\rm cdm}:\rho_{\rm b}\sim14:5:1$ today
calls for an explanation).
In a modified GR model, the way standard matter generates
potential wells or affect the dynamics of the universe is
changed. Note however that the new degree of freedom
are also gravitating so that in some models the distinction
is even more subtle.

Among the most studied alternative theories of gravity, let us
mention scalar-tensor theories discussed above, $f(R)$ theories
which are, after a field redefinition, a sub-class of scalar-tensor
theories, the DGP model~\cite{DGP} and the TeVeS~\cite{TeVeS} theory which is
a relativistic version of the MOND~\cite{milgrom} idea.

\subsection{Lessons for extending the tests to astrophysical scales}

GR is a well-defined theory of gravity with clear predictions so
that the consistency of these predictions offers the possibility
to test the theory in a model-independent way. This
implies that we need various observables relating the
same physical quantities (such as the mass in the example of the introduction).

In cosmology, we can use almost the same observations as
in the Solar system. Concerning light deflection, it
cannot be measured (since the ``undeflected'' position of the sources
cannot be determined; but for the particular case of microlensing) and we will have to use the distortion
of light bundles, that is strong and weak lensing. Also,
and contrary to the Solar system, we can have access
of the evolution of the energy of the photons, related
to the time variations of the gravitational potential
in the case of the integrated Sachs-Wolfe effect. The dynamics
of massive bodies can be obtained from the
large scale structure of the universe, which give
an information of the growth of the structures and their
velocity. Among the tests of the equivalence
principle, only the test on the constancy of fundamental
constants can be generalized.

There are however limitations specific to cosmology. In particular,
the cosmological structures evolve with time and this contains an
information on gravity but also on the properties of matter which
are difficult to disantangle (for instance, our prediction on the
shape of the galaxy power spectrum are different whether
there exist massive neutrinos or not). This also means that we
may have to take evolution effects into account. 
Also, cosmological data have to be interpreted in a statistical
way so that we always have a dependence of the initial
conditions that cannot be forgotten. Then, the description
of the dynamics of the universe involves the Copernican
principle so that the interpretation of our tests will
depend on such a hypothesis.

It follows that the tests that will be designed are
indeed tests of GR but also depends
on many other hypothesis so that they should probably be
considered first as tests of the $\Lambda$CDM model.

\section{Galaxy and Cluster scales}\label{sec:3}

The first interesting systems for testing GR in astrophysics are
galaxies and clusters. It is now well-established that, as
long as one assumes GR to hold, their dynamics can only
be understood by invoking the existence of dark matter.

The visible mass of spiral galaxies is rather concentrated so that
Newtonian gravity predicts that the rotation curves should drop as $r^{-1/2}$
outside the bright part of these galaxies. But this has not been confirmed by
more than a hundred rotation curve measurements~\cite{dm1}. Actually,
in most spiral galaxies, and more particularly those with a high surface brightness,
the rotation curves flatten at large distance from the center, $v_\infty\rightarrow$~const.
Moreover, this velocity is correlated to the luminosity of the galaxy.
This correlation, known as the Tully-Fisher law, states that the luminosity
of the galaxy scales as  $v_\infty^4$, so that one expects that $v_\infty^4\propto M$,
for the total stellar mass. This has provided the basis of the dark matter 
explanation: if the velocity is constant in the outer region of the galaxy, this
means that the centripetal acceleration scales as $a_r\propto r^{-1}$ and
Newton's law implies that the gravitational potential scales as $\ln r$.
In the case of  spherical symmetry, the Poisson equation implies that it
should be sourced by a matter whose density profile scales as $\rho(r)\propto r^{-2}$,
as for an isothermal sphere model. Thus, each spiral galaxy must
contain a spherical dark matter halo with a density profile scaling
as $r^{-2}$ at large distance. This reflects the discrepancy between two
estimations of the mass: the luminous mass and the dynamical mass.

To avoid such an hypothesis, Milgrom~\cite{milgrom} proposed
a phenomenological modification, called MOND, that was able to account for the 
galaxy rotation curves~\cite{sanders}, and more particularly
to recover the Tully-Fisher law. MOND introduces
a fundamental acceleration $a_0$, of the
order of $1.2\times10^{-10}{\rm m}\cdot{\rm s}^{-2}$,  such that the acceleration of any
massive body is
$$
 a = a_N,\quad\hbox{if}\quad a>a_0,\qquad
  a = \sqrt{a_Na_0},\quad\hbox{if}\quad a<a_0 
$$
so that, at large distance, the Newtonian acceleration being $GM/r^2$,
the centripetal acceleration is $\sqrt{GMa_0}/r$. Since it is also
given by $v^2/r$, one deduces that $v(r)\rightarrow(GMa_0)^{1/4}$. In this
regime, the gravitational potential behaves as $\sqrt{GMa_0}\ln r$ instead
of the standard Newtonian potential $-GM/r$. It follows that the deflection
angle at large distance from the center of the galaxy is
$$
 \Delta\theta_{\rm MOND} = \frac{2\pi\sqrt{GMa_0}}{c^2}.
$$
This value is the same as the one expected from GR, as
long as one is in the halo. Indeed, if interpreted within GR,
the presence of dark matter suggested by the rotation curves is
confirmed by lensing observations. Therefore in MOND,
an in any modification of GR, one must predict that a given
mass generates a larger potential and a larger deflection angle than 
in GR.

It follows that, one needs to estimate the mass of the galaxy,
in a given theory of gravity.
by different methods in order to check their compatibility. In particular,
different notions of mass needs to be distinguished~\cite{gefbruneton}: the {\it baryonic mass}
$M_{\rm b}$ assumed to be proportional to the luminous mass, or
stellar mass $M_*$; the total {\it dynamical mass} $M^{\rm dyn}_{\rm tot}$,
estimated from the rotation curves; and the total {\it lensing mass} 
$M^{\rm dyn}_{\rm tot}$ determined
by lensing observations. In the dark matter interpretation, and as well
established by lensing observations, we have
$$
\hbox{Dark Matter}:\qquad  M_{\rm b} <  M^{\rm dyn}_{\rm tot}\sim M^{\rm lens}_{\rm tot}.
$$
In particular, such mass estimates were performed in Ref.~\cite{if1} using
six strong lensing galaxies from the CASTLES database. The total mass
was estimated from lensing while the stellar mass was estimated from
a comparison of photometry and stellar population synthesis.
It demonstrates
that dark matter is still needed  (and that it is detected even in region
where $a>a_0$). In particular this dark matter component cannot be
explained by 2~eV neutrinos (see below).

If one assumes that the light deflection is given as in GR, then the previous
equivalences told us that MOND predicts the same lensing as in GR
within the dark matter halo. In particular the convergence at distance $r$
is given by $\kappa(r)=r_E/2r$, where the Einstein radius is
$$
 r_E = 2\pi\left(\frac{v_\infty}{c}\right)^2\frac{D_{ls}}{D_s}\quad \hbox{(MOND)}
 \qquad
  r_E = 4\pi\left(\frac{\sigma_v}{c}\right)^2\frac{D_{ls}}{D_s}\quad \hbox{(DM)},
$$
where the latter holds for a singular isothermal sphere with line-of-sight
velcity dispertion $\sigma_v$. While formally similar, these expressions
have however an interesting difference~\cite{tian} since $\sigma_v^2$
scales as $ M_{\rm tot}$ while $v_\infty^2$ scales as $\sqrt{M_*}$ so that
$$
 r_E \propto \sqrt{M_*}\quad \hbox{(MOND)}
 \qquad
  r_E \propto M\quad \hbox{(DM)}.
$$
The scaling of the Einstein radius with the stellar mass was measured~\cite{tian} 
using the RCS and SDSS surveys to show that $r_E\sim M_*^{0.74\pm0.08}$.
This seems in contradiction with the MOND prediction but the data 
used measurements of the shear at distances of some hundred
of kpc, at which the environment effects can change the MOND prediction.
It sets no constraint and the cold dark matter model since the fraction
$M_*/M$ is not known.  

Indeed, MOND is a phenomenological description but not a field theory.
As, we have seen earlier, the light deflection in scalar-tensor theories
in smaller than in GR. This means that a MOND cannot derive from
a simple scalar-tensor theory. It was realized (see Ref.~\cite{gefbruneton}
for more details) that this can be solved by coupling matter not to
the metric $g_{\mu\nu}$ but rather to a ``physical metric'' involving both a scalar
and a vector field
$$
 \tilde g_{\mu\nu} = \hbox{e}^{-2\phi}g_{\mu\nu} - 2U_\mu U_\nu\sinh2\phi.
$$
The first term is similar to what is performed in scalar-tensor theories and
the second term, involving the vector field $U_\mu$, allows to
reconcile light deflection with the GR prediction. This theory
is known as the TeVeS (Tensor-Vector-Scalar) theory~\cite{TeVeS}. 
Ref.~\cite{zhaomond} compares the TeVeS predictions to a large sample of galaxy
strong lenses from the CASTLES sample. Recently, Ref.~\cite{if2} compares
the predictions of TeVeS for both galaxy rotation curves and strong lensing
(for high and low surface brightness galaxies)
concluding that TeVes, in its simplest form, cannot reproduce these
data consistenly. The analysis~\cite{schwab} of galaxy-scale strong lensing from
the Sloan ACS (SLACS) survey indicates that $\vert\gamma^\ppn-1\vert<5\times10^{-2}$.
However this work emphasizes that setting constraints with such
systems requires to know the properties of the lensing galaxies with a great accuracy,
much greater than at present. The comparison of the stellar velocity dispertion
to measurements of the Einstein radius allows to constrain $\gamma^\ppn$,
reaching~\cite{smithkpc}  $\gamma^\ppn=0.88\pm0.05$ on kiloparsec scales
at 68\% C.L. while an early analysis based on 14 systems only gave~\cite{bolton}
$\gamma^\ppn=0.93\pm0.1$.

On cluster scales, various estimates of the mass can be obtained by lensing
(strong and weak), X-ray emission that characterizes the intracluster
(baryonic) medium, and the SZ effect which gives an information of the electron
distribution. By comparing these distributions, one can compare the location
of the gas and the gravitational iso-potential. Earlier analysis used the comparison
of X-ray and strong lensing~\cite{miralda} and then weak lensing~\cite{squires}
leading to the conclusion that the Poisson equation should be valid, within a factor 2, up
to scales of 2~Mpc~\cite{allen}.

Interesting conclusions arise from the study of the colliding galaxy
clusters 1E0657-56 ($z=0.296$). In this system a smaller cluster, known as the  ``bullet cluster'',
has crashed through a larger one and their intracluster gas has been stripped
by the collision. On one hand, weak lensing shows that the lensing mass
is concentrated in the two regions containing the galaxies rather than in
the stripes containing the baryonic matter~\cite{clowe}. A similar
observation~\cite{bradac2} was made with the merging galaxy cluster
MACS~J0025.4-1222 ($z=0.586$) for which the emitting gas, traced 
from its X-emission, is clearly displaced from the distribution of galaxies (from
lensing). The rich cluster Abell 520 ($z=0.201$) also exhibits the same
properties~\cite{abel1} and contains a massive dark core, as deduced from lensing mass reconstruction,
that coincides with the central X-ray emission peak. 
The analysis~\cite{abel2} of the cluster Abel~478 demonstrates that the X-ray, SZ and weak lensing
data perfectly agree with a dark matter model (but does not prove they cannot
be reproduced by a MOND model).

This seems to be a proof of the need
of dark matter since, being collisionless, it continues to be located around the
bullet, contrary to the baryonic gas. This was confirmed~\cite{bradac1} by
the reconstruction of the mass distribution from both strong and weak lensing.
However, it seems that MOND could accomodate these observations in particular
because the original TeVeS versions make different prediction when the system
is not spherically symmetric. Ref.~\cite{angus1} showed that it was possible
to design a multi-centred baryonic system eproducing
the weak-lensing signal of 1E0657-56  with a buller-like light
distribution. The same authors~\cite{angus2} then realized that a purely baryonic
MOND model cannot accomodate the data and that the bullet cluster was
dominated by dark matter whether one uses GR or MOND. In the latter
case, it would require massive neutrinos with $m_\nu=2$~eV.

The existence of such neutrinos seems however problematic. From
the study weak lensing for 3 Abell clusters and 42 SDSS clusters, it
was concluded that MOND cannot explain the data unless some
dark matter is added and this dark matter cannot be accounted for
by massive neutrinos~\cite{chibamond}. This was confirmed~\cite{priya} by the
confrontation of strong and weak lensing from the HST Wide-Field
Camera, excluding the dark matter to be neutrinos with mass in the range 2-7~eV.

In conclusion, it seems that MOND and TeVeS have difficulties to reproduce
the observations of the distribution of dynamical, baryonic and lensing
masses. Indeed none of the above mentioned results demonstrate that
MOND is ruled out. They are analysis that show that the data
can be consistently interpreted assuming GR and the existence 
and dark matter.  One of the main difficulty to use these observations
as a direct test of GR is the complex geometry of the systems
that are used.

\section{Cosmological scales}\label{sec:4}

The construction of a cosmological model relies on the choice of the theory of gravity as well
as on our understanding of the fundamental interactions of nature. However it also involves
other hypothesis, such as the Copernican principle which states
that we are not seating in a priviledged place of space. Under such
an hypothesis, and whatever the theory of gravity, the universe on large scales
can be described by a Friedmann-Lema\^{\i}tre spacetime with
metric
\begin{equation}\label{FLmetric}
 \dd s^2 =-\dd t^2 + a^2(t)\gamma_{ij}\dd x^i\dd x^j,
\end{equation}
where $t$ is the cosmic time, $a$ the scale factor and $\gamma_{ij}$ the spatial metric
on constant $t$ hypersurfaces. If one assumes that
GR is a good description of gravity then the dynamical equations of such a model derives
from the Einstein and the conservation equations
\begin{equation}
 G_{\mu\nu} = 8\pi G T_{\mu\nu}, \qquad
 \nabla_\mu T^{\mu\nu} = 0,
\end{equation}
where $ G_{\mu\nu}$ is the Einstein tenor and $T_{\mu\nu}$ the total stress-energy
tensor. Among this class of models our reference model is the $\Lambda$CDM model
which includes a cosmological constant and cold dark matter and assume that initial
conditions are consistent with the prediction of slow-roll inflation. Such a model is
in very good agreement with most of the astrophysical data and
it is self-consistent. But the fact that
the dark sector represents 95\% of the matter content of the universe 
and the cosmological constant problem drive us to test the hypothesis of our model,
and in the first place the Copernican principle and GR. It is indeed difficult to
anbandon these two hypothesis at the same time so that all the studies
aiming at testing GR in that context assume that the spacetime metric
remains of the form~(\ref{FLmetric}). Also most of them still include
dark matter and aim at replacing dark energy by a modification of GR.

Two roads can be followed. Either one defines a class of gravity models
that contains GR in some limit and then confronts it to cosmological data
to see how close from GR, in this particular space of theories, the theory
of gravity should seat. As an example, this was performed in depth
for scalar-tensor theories~(\ref{STaction}) for which the implications
of the background dynamics~\cite{ru1}, of the cosmic microwave background~\cite{ru1},
primordial nucleosynthesis~\cite{couv1}, weak lensing~\cite{sur04}, 
and local constraints~\cite{msu}, even allowing
for extensions to a non-universal coupling of dark matter~\cite{couv2}, were
all studied. Or one tries to quantify the allowed deviations from the
reference model while being as much as can be model-independent.
The strategy is then to exhibit consistency relations, analogous to
Eq.~(\ref{relSS}),
betwen different observables, which must hold in our
$\Lambda$CDM reference model, as first
proposed on the particular case of the Poisson
equation in  Ref.~\cite{ub01}. As explained in Ref.~\cite{uzanGRG06}, 
the modification of our reference framework can be classified in universaltiy
classes who have specific signatures and different tests can favour
or disfavour some classes of modification.

\subsection{Background dynamics}

The dynamics of the background spacetime is dictated by the
Friedmann equations
\begin{equation}\label{FLeq}
 H^2 = \frac{8\pi G}{3}\rho - \frac{K}{a^2} + \frac{\Lambda}{3},
 \qquad
 \frac{\ddot a}{a}= -\frac{4\pi G}{3}(\rho+3P)  + \frac{\Lambda}{3}
\end{equation}
where $H=\dot a/a$ is the Hubble function and $K=0,\pm1$ is the
curvature of the spatial sections. From the last equation, the recent acceleration of the cosmic 
expansion implies that $(\rho+3P)<0$ if GR is a good description of gravity.

At the background level, a modification of GR 
or the introduction of a dark energy component instead of the cosmological
constant will change the Friedmann equation and can be taken into account
in a effective way simply in terms on an effective fluid
\begin{equation}\label{FLeqmodif}
 H^2 = \frac{8\pi G}{3}(\rho + \rho_\dark) - \frac{K}{a^2} ,
 \qquad
 \frac{\ddot a}{a}= -\frac{4\pi G}{3}(\rho+3P+\rho_\dark+ 3 P_\dark),
\end{equation}
where $\rho_\dark$ and $P_\dark$ can depend on $H$ and its derivatives,
as e.g. for scalar-tensor theories or DGP. This allows to define the equation of state
of the dark energy from $H$ (see Ref.~\cite{msu}).

Indeed, without an explicit model, the extra-terms $\rho_\de$ and $P_\de$
are not known. Besides we known that, since they arise from the existence
of a new degree of freedom, there must exist an associated equation of 
evolution~\cite{ueamphy}. The standard approach is to postulate that
they are related by an equation of state, $\rho_\de=w_\de P_\de$, and the
most commonly used ansatz is~\cite{cp,linderw}
\begin{equation}\label{w_cpl}
 w_\de(z) = w_0 +\frac{z}{1+z} w_a,
\end{equation}
with $w_0$ and $w_a$ constant. It is thus clear that the background
dynamics cannot distinguish between a modification of GR and
a properly tuned dark energy model. This lies in the fact that
the only quantity at hand is $H(z)$ and most of the models
of the literature can be tuned to reproduce the same function
(see Ref.~\cite{uzanGRG06} for explicit examples).
No null test of GR can be constructed with background data since they
all are functions of $H(z)$.

\subsection{Linear perturbation theory}

\subsubsection{Standard $\Lambda$CDM}

As long as one assumes GR to hold and consider an almost
Friedmann-Lema\^{\i}tre spacetime, the evolution of perturbations
is well understood, see e.g. Ref.~\cite{pubook}. 
On sub-Hubble scales, focusing only on scalar perturbations 
which are dominant at late time, the space-time metric can be written as
\begin{equation}\label{FLpertmetric}
 \dd s^2 =-(1+2\Phi)\dd t^2 + (1-2\Psi)a^2(t)\gamma_{ij}\dd x^i\dd x^j,
\end{equation}
where $\Phi$ and $\Psi$ are the two gravitational potentials.

The evolution equations on Hubble scales are given by the conservation
of the matter stress-energy tensor (continuity and Euler equations)
\begin{equation}\label{pert1}
 \dot\delta_m=-\frac{\theta_\mat}{a}, 
 \qquad
 \dot\theta_\mat+H\theta_\mat=-\frac{1}{a}\Delta\Phi,
\end{equation}
which leads to the standard second order evolution equation
\begin{equation}\label{pert2}
 \ddot\delta_\mat +2H\dot\delta_\mat -\frac{1}{a^2}\Delta\Phi = 0.
\end{equation}
Among the Einstein equations, we can keep only the Poisson equation
\begin{equation}\label{pert3}
 \Delta\Psi = 4\pi G\rho_\mat a^2\delta_\mat
\end{equation}
and
\begin{equation}\label{pert4}
 \Phi-\Psi = 0
\end{equation}
that arises from the fact that the matter anisotropic stress is negligible. This shows
that the two gravitational potentials have to coincide, which is related to the
fact that $\gamma^\ppn=0$ in GR, their spectrum has to be proportional
to the matter power spectrum, as firt pointed out in Ref.~\cite{ub01} and Eq.~(\ref{pert1})
implies that
\begin{equation}
 \theta_\mat =-f \delta_\mat,
\qquad
\hbox{with}\qquad
 f = \frac{\dd\ln D}{\dd\ln a},
\end{equation}
where we have decomposed the density contrast as $\delta_\mat = D(t)\epsilon(x)$ where $\epsilon$
encodes the initial conditions. It was shown~\cite{lahav91} that,
for a flat CDM model,
\begin{equation}
 f \sim \Omega_\mat^{0.6}
\end{equation}
was a good fit. The important feature here is that if GR holds, then the growth of
structures is completely determined by $H(z)$. We forsee that one can
check the compatibility of
background data (such a distance-redshift relations, e.g.
from SNIa) and large scale structure data, whatever the
parameters entering the equation of state~(\ref{w_cpl}) or
any other parameterisation (see below).

\subsubsection{Extension}

In order to construct tests of GR in this regime, one needs to construct
the most general extension of this set of evolution equations, still
assuming we are dealing with a metric theory of gravity. 

In the case of dark energy alone, one needs to consider the effect of its stress-energy tensor,
which can have non-vanishing anisotropic stress and density contrast, contrary
to a pure cosmological constant but the equations of evolution of the other fluids
are not modified, since otherwise this new component
would be coupled to the standard matter non minimally. 
In GR modifications, there exists a new long range force and
the evolution equation of matter will be of the form
$\nabla_\mu T_i^{\mu\nu} = f_i^\nu$,
where $f^\mu$ is a force term between the standard matter
fields and the new degree of freedom. The way such force term
appears in the equation and its relation to the
Einstein equation is not obvious to describe in full generality while
being model-independent. As an example, consider scalar-tensor
theories of gravity. In the Jordan frame, the equations of motion
of the standard matter fields are not modified so that $f_i^\mu=0$ but,
performing a conformal transformation, the same theory, written in the
Einstein frame, involves a force $f^i_\mu = \alpha(\varphi_*) T_i\partial_\mu\varphi_*$
that will appear even at the background level in the continuity equation. In this
particular case, it is well understood that the modification of gravity appears
as a time-dependent modification of the Newton constant in the Jordan frame
while it is seen as a universal time-dependent modification of masses in the Einstein frame.
Indeed, if the new force is not universal, it probably involves that mass ratios
will be time-dependent, which can be tested~\cite{uzan03}. Thus, we assume that
we are working in the equivalent of the Jordan frame so that we assume
that $f_i^0=0$ and that there is no creation of matter. The spatial
component of the force can however be non vanishing and enters the Euler equation. 
Let us stress that while important from a physical point of view~\cite{cqgjpu,kms}, this is not
dramatic from a phenomenological point of view since only the source term
of Eq.~(\ref{pert2b}) below will be changed. On the other hand, we
shall have an equation of evolution for the new degree of freedom that shall
also have a source term proportional to the matter stress-energy tensor~\cite{cqgjpu}. 
Unfortunately, this equation remains unkown until we specify the model.

As long as we stick to linear perturbations, these extensions can be
implemented by modifying the previous equations as
\begin{equation}\label{pert1b}
 \dot\delta_m=-\frac{\theta_\mat}{a}, 
 \qquad
 \dot\theta_\mat+H\theta_\mat=-\frac{1}{a}\Delta\Phi + {\cal S}_\de,
\end{equation}
which leads to the standard second order evolution equation
\begin{equation}\label{pert2b}
 \ddot\delta_\mat +2H\dot\delta_\mat -\frac{1}{a^2}\Delta\Phi = {\cal S}_\de.
\end{equation}
The term ${\cal S}_\de(k,a)$ encodes the new long-range force between
the new degree of freedom and the standar matter
(and dark matter!).

Then, we need to write down the Einstein equations. First, we can
generalize the Poisson equation, written in Fourier space, as
\begin{equation}\label{pert3b}
  -k^2\Psi = 4\pi GF(k,H)\rho_\mat a^2\delta_\mat + \Delta_\de.
\end{equation}
The first term $F(k,H)$ accounts for a scale dependence of the
gravitational interaction while $\Delta_\de$ accounts for
a possible clustering of the new degree of freedom, and
in particular of dark energy if there is no modification of GR
(this shows at this stage, that the Poisson equation can be
modified without modification of GR if dark energy can
cluster; also care needs to be taken in the case of massive
neutrinos which can enter on the r.h.s. of this equation,
see e.g.~\cite{mnu}; Ref.~\cite{kunz06,jain07}
proposed a
interesting example of a clustering dark energy model
mimicking the DGP model).
Then, there is the possibility to have an effective anisotropic stress
so that
\begin{equation}\label{pert4b}
 \Delta(\Phi-\Psi) = \Pi_\de.
\end{equation}
It follows that the deviation from GR is encoded in the four functions
$({\cal S}_\de,F,$ $\Delta_\de,\Pi_\de)$ which, in the
case of the $\Lambda$CDM model, reduces to $(0,1,0,0)$ and,
in the case of dark energy to $(0,1,\Delta_\de,\Pi_\de)$, even though
in most cases $\Delta_\de$ and $\Pi_\de$ are negligible. Their
expression for quintessence, scalar-tensor and DGP models
can be found in Ref.~\cite{uzanGRG06}. For the same reason that, in the case
of a GR modification, at the background level $P_\de$ and $\rho_\de$ can depend on
$H$, ${\cal S}_\de$, $\Delta_\de$ and $\Pi_\de$ can depend on $\Phi$ and $\Psi$,
while $F$ is a function of the background quantities only.

Equations~(\ref{pert1b}-\ref{pert3b}) imply that the matter density
evolves as
\begin{equation}\label{growthgen}
 \ddot\delta_\mat+2H\dot\delta_\mat -4\pi G F(k,a)\rho_\mat a^2\delta_\mat
  = \mathcal{C}_\de
\end{equation}
with $\mathcal{C}_\de=(\Delta_\de+\Pi_\de)/a^2-\mathcal{S}_\de/a$ so that
the matter power spectrum is expected to be deformed in shape, mainly because
of the $k$-dependence arising from $F$ and from the source term $\mathcal{C}_\de$.

This idea to construct such a post-$\Lambda$CDM parameterisation on sub-Hubble scales was first proposed in Ref.~\cite{uzanGRG06}, following the analysis of the particular case of scalar-tensor theory
by Ref.~\cite{sur04}. Several similar approaches were then designed in 
Refs~\cite{jain07,amendola07,hu07,zukin,mustafa07,caldwell07,daniel08,thomas08,songdore08,song08a,dent}
which went further in determining the relations with observational data
(see discussion below).
In particular Ref.~\cite{amendola07} uses, instead of $\Pi_\de$, a parameter $\eta$ defined
by 
$$\Phi=(1+\eta)\Psi,$$
so that it is a generalisation of the post-Newtonian 
parameter $\gamma^\ppn$, as first proposed in the particular case of scalar-tensor theories
in Ref.~\cite{sur04}. ($\Phi\not=\Psi$ in the Solar system is an indication
of a modification of GR because one is dealing with vacuum solution of
Einstein equation; again, this is not the case in cosmology since dark
energy can have an anisotropic stress).
Instead of using the the set $(f,\Delta_\de)$, most analysis,
including Refs.~\cite{jain07,amendola07,thomas08,songdore08,song08a}, assume
that the Poisson equation is modified to 
$$
-k^2\Psi = 4\pi G Q(k,a)\rho_\mat a^2\delta_\mat
$$
involving only one new function $Q$. In most cases at hand, this is a good approximation,
but in full generality we shoud distinguish the large-scale modification of GR and
the clustering of possible new degrees of freedom (see e.g. Ref.~\cite{kunz06,jain07}). 
In particular $\Delta_\de$
could be an independent random variable, not proportional to $\delta_\mat$. 
The limit $\Delta_\de=0$ corresponds to pure modification of GR with negligible
effect of the new degree to the total stress-energy tensor while $F=1,\Delta_\de\not=0$
corresponds to models of clustering dark energy (and may also incorporate the
effect of massive neutrinos; see e.g.~\cite{takada} for a study of this degeneracy).

As such, this description is not complete since we have no equation to describe the
evolution of the new degrees of freedom. At the background level, this gap
is often filled by assuming a parameterisation of the dark energy
equation of state. As we shall see, two roads may be followed from this
point: either one parameterized the unkwon functions that appear here
or one construct null-tests.

\subsection{Cosmological data}

The previous analysis shows that in order to constrain the deviation from
GR with the large scale structure of the universe we need to be able to extract
information on the distribution of the four variables $(\Phi,\Psi,\delta_\mat,\theta_\mat)$
which are not directly observable. Let us summarize briefly some of the
observations that turn to be useful.  We call $P_{XY}(k,z)$ the 3-dimensional power
spectrum of the fields $X$ and $X$ at redshift $z$ (or equivalently time $t$ or distance
$\chi$) defined by $\langle X(\bk,z)Y(\bk',z)^*\rangle=(2\pi)^3P_{XY}(k,z)\delta^{(3)}(\bk-\bk')$
and $C_{XY}(\ell,z)$ their 2-dimensional (or angular) power spectra.

\subsubsection{Background data}

Background data usually include the luminosity distance-redshift relation
pro\-bed by SNIa which provides a handle on $H(z)$ up to $z\sim1.5$,
the angular diameter distance mainly from the tangential component
of the BAO measured by their imprint on galaxy distribution.

\subsubsection{Large scale structure}

The clustering of galaxies is one of the oldest measures
of the properties of the large scale structure. The galaxy power spectrum $P_{gg}$
is the simplest measure of the correlations in the galaxy number density $n_g$.
In general the distribution of galaxies is biased with respect
to the mass distribution and it is often assumed that they can be related
by
\begin{equation}
 \delta_g\equiv\frac{\delta n_g}{n_g}=b_1\delta_\mat + \frac{b_2}{2}\delta_\mat^2,
\end{equation}
where $b_1$ and $b_2$ are biais parameters. It is expected that
the biais is a complicated function of time
and of the masses of the halo hosting the galaxies~\cite{fbrev}, since it encodes in
some way all the process of galaxy formation. We shall assume here,
for simplicity, that we restrict to a linear biais so that
$P_{gg}(k,z)=b_1^2(z)P_{\delta_\mat\delta_\mat}(k,z)$.
Imaging data with photometric redshift provides a measurement of
the angular power spectrum of galaxies, which is a simple projection
of the three-dimensional power spectrum
\begin{equation}
 C_{gg}(\ell) = \int \frac{W^2_g(\chi)}{S_K(\chi)^2}P_{gg}\left(k=\frac{\ell}{S_K(\chi)},\chi\right)
 \dd \chi,
\end{equation}
$S_K$ being the comoving angular distance and where $W_g$ is the normalized redshift
distribution of galaxies in the sample.

However, the redshift-space position of any galaxy differs from its real
space position due to its peculiar velocity. The density contrast
in redshift space, $\delta^s_g$, and in real space, $\delta_g$, can be related by imposing
mass conservation~\cite{kaiser87}. In the linear regime, this leads to
$\delta_g^s=\delta_g +\mu^2\theta_g/aH$, where $\mu$ is the cosine of the
line-of-sight angle so that the redshift-space power spectrum is
\begin{equation}
 P_{gg}^s(k,\mu)= P_{gg}(k) + 2\frac{\mu^2}{aH}P_{g\theta_g}(k) + \frac{\mu^4}{a^2H^2}P_{\theta_g\theta_g}(k).
\end{equation}
It is thus commonly modelled as~\cite{fbrev,kaiser87}
\begin{equation}
 P_{gg}^s(k,\mu)=\left[ P_{gg}(k) + 2\frac{\mu^2}{aH}P_{g\theta_g}(k) + \frac{\mu^4}{a^2H^2}P_{\theta_g\theta_g}(k)\right]F\left(\frac{k^2\mu^2\sigma_v^2}{H^2(z)}\right),
\end{equation}
where $F$ is a smoothing function and $\sigma_v$ is the 1-dimensional velocity
dispersion. The angular dependence enables to separate the different
components~\cite{percivalw} to get a measurement of the three spectra
$P_{gg}(k,z)$, $P_{g\theta_g}(k,z)$ and $P_{\theta_g\theta_g}(k,z)$, in particular
from the SDSS~\cite{SDSS} and 2dF~\cite{2dF} galaxy redshift surveys. These
low redshift analysis were extended to $z\sim1$ in Ref.~\cite{guzzo} using
the VIMOS-VLT Deep Survey (VVDS)~\cite{vvds}.

Indeed, in the standard $\Lambda$CDM, and in the linear regime, we
have that $\theta_g=-a\dot\delta_\mat$ so that its growth rate is $D_{\theta_g}=-a\dot D$
so that $\theta_g=aHf\delta_\mat$. In that limit  $P_{g\theta_g}=aH\beta P_{gg}$
and $P_{\theta_g\theta_g}=a^2H^2\beta^2P_{gg}$ with $\beta=f/b$, and
the three spectra are not independent.

\subsubsection{Weak lensing}

Gravitational lensing offers various posibilities. As previously,
we restrict our analysis to metric theories of gravity.

First, either in the strong or weak regime, it can probe the sum of the two Bardeen
potentials, $\Phi+\Psi$. Weak lensing surveys use the observed ellipticities
of backgroung galaxies (and more particularly the correlation of their shapes)
to reconstruct a map of the cosmic shear, which can then be used to 
determine the convergence $\kappa$~\cite{ymrev}. As long as
photons travel on null geodesics and the geodesic deviation equation
holds, the distortion of the shape of background galaxies can be
computed from the Sachs equation~\cite{sachs61} leading to
\begin{equation}
 \kappa(\bt,\chi_i)=\frac{1}{2}\int W(\chi,\chi_i) \Delta_2(\Phi+\Psi),
 \dd\chi
\end{equation}
for sources located in a bin centered round a redshift $z_i$ and with $\chi_i=\chi(z_i)$
with
$$
W(\chi,\chi_i)=\frac{S_K(\chi)S_K(\chi_i-\chi)}{S_K(\chi_i)}.
$$
The convergence
power spectrum for two sets of galaxies centered around $z_i$ and $z_j$,
as can be obtained by a tomographic survey, is thus
\begin{equation}
 C_{\kappa\kappa}(\ell,z_i,z_j)=\frac{\ell^4}{4}\int W(\chi,\chi_i)W(\chi,\chi_j)
 P_{\Phi+\Psi}\left(k=\frac{\ell}{S_K(\chi)},\chi\right)
 \dd\chi.
\end{equation}
Until we have data allowing for the use of tomography, we have only
access to the shear power spectrum averaged on the source redshift 
distribution, $P_{\kappa\kappa}(\ell)$ which is given by the same expression
but with the window function
$$
W(\chi)=S_K(\chi)\int_\chi W_g(\chi')\frac{S_K(\chi'-\chi)}{S_K(\chi')}\dd\chi'.
$$ 
In conclusion, this
allows to constrain the power spectrum $P_{\Phi+\Psi}(k,z)$.
Note that the analysis of weak lensing requires in fact to know
the non-linear power spectrum but the latest data from the
CFHTLS~\cite{fuwl} reach large angular scales ($\theta>30$ arcmin.), 
which allows  to work in the (quasi) linear regime, where theoretical
predictions for the time evolution of the power spectrum are more reliable.
Note also that the convergence power spectrum is often
expressed in terms of the matter power spectrum, making
use both of the Friedmann and Poisson equations. Indeed,
the goal here is to relate the observables to their primary perturbation
variables without using any equations.

Second, one can use galaxy-galaxy lensing, which arises when the deflecting and source
galaxies are aligned, giving rise to a mean tangential shear around foreground galaxies.
This will thus give an information on the correlation between the galaxy
distribution and $\Phi+\Psi$ through the angular power spectrum
\begin{equation}
 C_{g\kappa}(\ell,z_i,z_j)=\frac{\ell^2}{2}\int W_g(\chi,\chi_i)W(\chi,\chi_j)
 P_{g,\Phi+\Psi}\left(k=\frac{\ell}{S_K(\chi)},\chi\right)
 \dd\chi.
\end{equation}
This was for instance measured from the SDSS galaxy survey. Note
that the magnification biais~\cite{fbrev} can also help to extract some correlations since
$$
 \delta_g = b\delta_\mat +(5s-2)\kappa
$$
where $s=\dd{\rm log}N/\dd m$ is the logarythmic slope
of the number count-magnitude function. This induces
distortions~\cite{alberto,loverde} that can also be used to test GR.
Note that it was also proposed that the weak lensing of
standard candles (SNIa, or GW sirens) can be used to 
measure the cross correlation between the magnification $\mu$
and $\delta_g$~\cite{corray,lindergw}. In particular
$C^{\mu\mu}$ and $C^{g\mu}$ contain informations
similar to $C^{\kappa\kappa}$ and $C^{g\kappa}$ respectively.
This has not been investigated yet.

\subsubsection{Integrated Sachs-Wolfe effect}

The observation of the cosmic microwave background temperature
anisotropies gives numerous important informations for our
cosmological model, among which the initial power spectrum for
the perturbations. While propagating from the last scattering
surface to us, the energy of the photons changes due to the fact
that they cross structures in formation, and thus propagate
in a spacetime where $\Phi$ and $\Psi$ are not constant. This
induces a direction-dependent temperature change, known as
the integrated Sachs-Wolfe effect~\cite{pubook}
$$
 \frac{\Delta T}{T} = \int (\dot\Phi+\dot\Psi) a(\chi)\dd\chi,
$$
where the integral is performed along the photon geodesic. This
ISW effect is correlated with the galaxy distribution with angular
power spectrum
\begin{equation}
 C_{g,ISW}(\ell) =  \int P_{g,(\dot\Phi+\dot\Psi)}\left(k=\frac{\ell}{S_K(\chi)},\chi\right)
  \frac{a^2(\chi)}{\chi^2}\dd\chi.
\end{equation}
This has been detected~\cite{isw} by cross-correlating the CMB
anisotropies to galaxy maps.

\subsubsection{Conclusions}

We see that astrophysical observations allow to measure many
correlations between the perturbation variables.
In order to construct tests
of GR, one needs to relate $\delta_g$ to $\delta_\mat$ and thus understand
the biais (and most importantly contrain its scale dependence). The example
given above are the most promising to implement the tests of GR
but, indeed, there exist many other ways to measure these quantities.

We also need to keep in mind that each of this method has its
own systematics and limits. We cannot discuss this issue here,
but it is central when actually deriving constraints.

\subsection{Growth of matter perturbations}

The first effect of a modification of GR is to change the growth of
density perturbation.  In the $\Lambda$CDM model, Eqs.~(\ref{pert1}-\ref{pert2}) imply that
the growth rate $D$ evolves as
\begin{equation}
 \ddot D +2H\dot  D -4\pi G\rho_\mat D  = 0.
\end{equation}
This equation can be recast in terms of $a$ as time variable~\cite{pubook} as
\begin{equation}\label{evoDLCDM}
 D'' +\left(\frac{\dd\ln H}{\dd a} +\frac{3}{a} \right)D'
 = \frac{3}{2}\frac{\Omega_{\mat0}}{a^5}D,
\end{equation}
from which it can be checked that $D=H$ is a solution so that the
growing mode can be obtained as
\begin{equation}
 D = \frac{5}{2}\frac{H}{H_0}\Omega_{\mat0}\int_0^a\frac{\dd u}{[u H(u)/H_0]^3}  , 
\end{equation}
which implies that if $H(z)$ is known from background observations, such
as SNIa, then $D(z)$ is fixed. There is a rigidity between the
expansion history and the growth rate. 

The growth rate can be parameterized phenomenologically as~\cite{lindera,linderb}
\begin{equation}\label{defg}
 \frac{\dd\ln D}{\dd\ln a} = \Omega_\mat^\gamma.
\end{equation}
Then, if GR is not modified, 
the index $\gamma$ can be computed once $H(z)$, or
equivalently the dark energy equation of state, is known and it was 
shown~\cite{linder2,ws98}  that
$$
 \gamma=0.55 + 0.05[1+w_\de(z=1)].
$$
While being a good test of dark-energy model with a smooth energy distribution, it is
not clear whether it can be considered as a test of GR. In particular, we can imagine
that dark energy has an anisotropic stress or is clustering while GR is not
modified (i.e. $F=1$, $\mathcal{S}_\de=0$, $\Pi_\de\not=0$, $\Delta_\de\not=0$
so that  $\mathcal{S}_\de\not=0$) so that one accomodate a value of
$\gamma$ by some properly designed model. To finish, such a parameterisation
is too restrictive since it does not include the scale-dependence that is
expected from the modification of GR (see however Ref.~\cite{dent}). The extensions to include the
super-Hubble regime were considered in Refs.~\cite{zukin,bert,dent}.
It was recently proposed~\cite{ksz}
that galaxy cluster velocities, measured from the kinetic SZ effect,
may allow for a measurement of $\gamma$.

Several studies concentrate on a pure modification of the Poisson equation
so that only the term $F(k,H)$ is modified. The effect of such a $k$-dependent term
on the power spectrum was first studied in Ref.~\cite{ub01} where the
function $F$ was assumed to reproduce the effect of higher-dimensional
gravity, as described at the time. In that particular case, where the only
parameter is the length scale $r_s$, it was shown~\cite{fb04} that
the cosmic shear~\cite{shear3} 3-point function implies that $r_s>2h^{-1}$Mpc.

Similar analysis in the case
of a Yukawa type modification of the gravitational potential,
$$
 \Phi(\br) = -G\int\dd^3\br'\frac{\rho(\br')}{\vert\br-\br' \vert}
\left[1+\alpha\left(1 - \hbox{e}^{-\vert\br-\br' \vert/\lambda}
\right)\right] 
$$
were then performed. With such a potential, the function $F(k,a)$
entering the Poisson equation is given by~\cite{sealfon,shirata1}
\begin{equation}
 F(k,a) = 1 + \alpha\frac{(a/k\lambda)^2}{1  + (a/k\lambda)^2}.
\end{equation} 
Such a modification causes the rate of growth to depend on $k$
so that the scale $\lambda$ shall have an imprint on
the power spectrum (see also Ref.~\cite{lue} for
a general argument on the shape dependence). Interestingly, assuming an Einstein-de Sitter
bacground cosmology, Eq.~(\ref{growthgen}) can be solved analytically
in terms of hypergeometric function~\cite{sealfon,shirata1} to give the
growing mode
$$
 \delta_+(k,a) = s\,
 {}_2F_1\left(\frac{5-\sqrt{25+24\alpha}}{8},\frac{5+\sqrt{25+24\alpha}}{8},
 \frac{9}{4};-s^2 \right)
$$
with $s\equiv a/k\lambda$.

Ref.~\cite{sealfon} considered the case of a Einstein-de Sitter background
and analyzed the SDSS and 2dFGRS data up to $k\sim0.15h/$Mpc,
leading respectively to the constraints $\alpha=0.025\pm1.7$ and
$\alpha=-0.35\pm0.9$ at a 1$\sigma$ level. A similar
analysis was performed in Ref.~\cite{shirata1} who used
the Peacock and Dodds procedure~\cite{PD} to describe the non-linear
power spectrum. The analysis of the SDSS data sets the constraints
$-0.5<\alpha<0.6$ (resp. $-0.8<\alpha<0.9$) for $\lambda=5h^{-1}$~Mpc
(resp. $\lambda=10h^{-1}$~Mpc). The analysis was extended in Ref.~\cite{shirata2}
by performing both second order perturbations and N-body simulations to
construct a mock galaxy catalog. Ref.~\cite{sereno} extended these analysis by allowing
a modified expansion rate, which should be the case if GR is modified.
They also showed that the modification of the shape of the power
spectrum is almost degenerate with the effect of massive neutrinos.
Notice that a combined analysis~\cite{martig} using CFHTLS weak lensing data and the
SDSS matter power spectrum estimated from lumianous red galaxies 
found no sign of deviation from GR on scales ranging between
0.04 and 10 Mpc. Even though this analysis used both matter distribution
and weak lensing, it only constrained the shape of the power
spectrum without implementing the consistency chek proposed
in Ref.~\cite{ub01}.

N-body simulations with such a Yukawa modification of the gravitational
potential were performed in Ref.~\cite{stabenau} who concluded
that the gravitational evolution is almost universal, at least for $\lambda$ in the $1-20$~Mpc
range so that the Peacock and Dodds approach~\cite{PD} can
be adapted to get an analytical fit. It was extended by the simulations of Ref.~\cite{laszlo} which include
the possibility of an anisotropic stress and considered the case of
DGP models with $r_s=(5,10,20)h^{-1}$~Mpc. To finish,
the spherical collapse model and the estimate of the abundance of
virialized objects was considered in Refs.~\cite{martino,HJ}.
The scale dependence of the growth rate was proposed~\cite{acqua08} to be studied
in terms of
$$
 \epsilon(k,a)= \Omega_\mat^{-\gamma}(a)\frac{\dd\ln D}{\dd\ln a} -1,
$$
which remains close to 0 for any smooth dark energy model. In particular
it can be measured from future redshift surveys.

These studies allows to understand the effect of the modification of
the Poisson equation, which is expected to be generic in any
deviations from GR, and absent in all models of pure dark energy.
They are thus very instructive but note that the background cosmology
is in general not modified in a consistent way.

\subsection{Testing GR on cosmological scales}

There have been two main approaches to using cosmological data
to constrain deviations from GR.

\subsubsection{Parameterizing our ignorance}

In the first approach, one tries to use the generalized set of
perturbation equations~(\ref{pert1b}-\ref{pert4b}) in order
to compute various cosmological observables and compare them
to astrophysical data. The main problem, as mentioned
earlier, is that such a parameterisation cannot be complete
unless the physics of the new degrees of freedom is know.
We thus have two possibilities. Either one compute explicitely
these terms in some classes of theories such as $f(R)$, DGP or 
scalar-tensor~\cite{uzanGRG06,amendola07,hu07} or one specifies
some ans\"atze for these functions, in the same spirit as we introduced the 
parameterisation~(\ref{w_cpl}) for the dark energy equation of state.

For instance, Refs.~\cite{amendola07,thomas08}
assume that the function $\Sigma\equiv Q(a,k)(1+\eta(a,k)/2)$
can be expanded as $\Sigma= 1+ \Sigma_0a$ with $\Sigma_0$
constant. The effect of the modification of GR is taken into account
through a parameterisation of the form~(\ref{defg}), with $\gamma$
constant so that one ends up with 4 constant extra-parameters $(w_0,w_a,\gamma,\Sigma_0)$ 
besides the standard cosmological parameters, the $\Lambda$CDM
model corresponding to $(w_0,w_a,\gamma,\Sigma_0)=(-1,0,0.55,0)$. Such a
parameterisation was then used to discuss the sensitivity of various probes. 
Clearly, this choice
misses a possible scale-dependence of $F$ (or $Q$) which is generically
expected if GR is modified~\cite{lue}. This issue was recently adressed
in Ref.~\cite{guzik09} which proposes
to expand the two unknwon functions $Q$ and $\eta$ as
$$
X(a,k)\simeq X_0(a)+X_1(a)aH/k.
$$ 
Ref.~\cite{zhao08} chooses
the same two functions as functions of $k$ and $a$ and
proposes different ans\"atze for their functional form in order
to study the potential of upcoming and future tomographic
surveys to constrain them. Ref.~\cite{hu07} proposed  a
parameterisation that depends on the scale.
On the other side Refs.~\cite{caldwell07,daniel08}
focus only on the function $\eta(z)$, that they call $\varpi(z)$, in order to
infer its influence on CMB anisotropy spectrum and
weak lensing. $\varpi$ was chosen to scale as $\varpi_0\rho_\de(z)/\rho_\mat(z)$,
assuming a $\Lambda$CDM evolution for the background cosmology.

Note that in the standard $\Lambda$CDM, we must have
that $\Psi(\bk,a)=\Phi(\bk,a)=-3\Omega_\mat(a)(Ha/k)^2\delta_\mat(\bk,a)/2$
and $\theta_\mat(\bk,a)=-f\delta_\mat(\bk,a)$. Thus, instead of parameterizing the
unknown terms that enter the perturbation equations, we may think
to parameterize directly their solution, for instance, as
$$
 \Phi(\bk,a)=-\frac{3}{2}\Omega_\mat(a)(Ha/k)^2\delta_\mat(\bk,a)\left[1 + c_\Phi(k,a)\right],
$$
$$
  \Psi(\bk,a)=-\frac{3}{2}\Omega_\mat(a)(Ha/k)^2\delta_\mat(\bk,a)\left[1 + c_\Psi(k,a)\right],
$$
and
$$ 
  \theta_\mat(\bk,a)=-f\delta_\mat(\bk,a)\left[1 + c_\theta(k,a)\right],
$$
where $\delta_\mat(\bk,a)$ is supposed to scale as
$\delta_\mat(\bk,a)=\delta_\mat(\bk,a_{in})D(a)\left[1 + c_\delta(k,a)\right]$
and then finding physically motivated ans\"atze for the functions $c_i$. 
Such an alternative parameterisation was considered in Ref.~\cite{mustafa07}.

In these approaches, the game is thus to replace free unkown functions
by a set of parameters in order to be able to compute the different
cosmological observable and then compare them to data. This allows in particular 
to understand the accuracy with which they can be constrained
by forthcoming experiments, the main difficulty being to find the most relevant set
of parameters that reproduce a large class of theories. Ref.~\cite{thomas08}
utilised the large angular scales ($\theta>30$ arcmin.) weak lensing data
from the CFHTLS (in order to work in the linear regime), BAO and SNIa data
to get constraints on $(\Omega_{\mat0},\Sigma_0,\gamma)$ consistent
(e.g. related to the parametrisation of the equation of state)
with the standard $\Lambda$CDM. Note also that it is important
that the background and perturbation dynamics be consistent since they
derive from the same modification of GR, an issue often overlooked.

\subsubsection{The art of correlating}

A probably  better idea to obtain constraints on deviation from
GR is to construct null tests. Such tests are based on the simple
fact that once the theory is completely specified, there
must exist consistency relations, between different observables,
in a similar way as the Solar system example of the introduction led
to the consistency relation~(\ref{relSS}). Indeed, any departure from
such a relation would indicate that some hypothesis of our
model are not correct and that the theory needs to be extended,
without telling how. Such tests are {\it null tests}, in the
spirit of ``traditional'' physics in which a reference model is
confronted to observations in order to determine the limits of its
validity. 

The use of cosmological data to perform such tests was first
proposed in Ref.~\cite{ub01} who focused on the Poisson
equation~(\ref{pert3}). If such an equation holds then the
power spectra of the gravitational potential $P_\Phi(k)$ and
of the matter distribution $P_{\delta_m}(k)$ must be related by
$$
 k^4 P_\Phi(k,a) = \frac{9}{4}\Omega_{\mat0}H_0^2 a^{-2}
 P_{\delta_\mat}(k,a),
$$
whatever the cosmological scenario. This means that the scale
dependence of the two pectra are related in a very specific way.
In particular, if the Poisson equation is modified, the change
of the shape of the matter power spectrum is, as we saw on the
example of a Yukawa potential above, model dependent but the fact
that the two spectra differ is a model independent conclusion.
In particular such a relation can be tested by comparing weak
lensing data to galaxy survey, if the scale dependence of the biais
is mild, as expected from numerical simulation, since $C_{\kappa\kappa}$
and $P_{gg}$ give access to $P_\Phi$ and $P_{\delta_\mat}$. Note also that, it has a trivial
generalisation if the fields are all proportional to the same
stochastic variable (which is the case for adiabatic initial
conditions) then $P_{\Phi\delta_\mat}=\sqrt{P_\Phi P_{\delta_\mat}}$,
which again can be tested using galaxy-galaxy lensing.

Similar rigidities were exhibited between the background dynamics
and the growth of the large scale structure. For instance, the
growth equation~(\ref{evoDLCDM}), valid for a $\Lambda$CDM,
and thus when GR hold, can be recast~\cite{chiba07,nesseris07}
(see also Refs.~\cite{linderz,ywang08})
as a first order equation for $H$ so that $H(z)$ can be inferred from background data 
and perturbation data independently, or equivalently the equation of state
of dark energy ~(\ref{w_cpl}) and the parameter $\gamma$ defined
in Eq.~(\ref{defg}) are not independent when the dark energy is assumed
to remain smoothly distributed.
This was implemented  in the analysis of Refs.~\cite{wang07,ishak06}
who introduce two dark energy equations of state, one for the evolution
of the background geometry and the other governing the growth.
Using SNLS-SNIa, 2dF and SDSS galaxy redshift survey, CMB data and CTIO-lensing
survey, they concluded that the two determinations of $\Omega_\Lambda$
were consistent and that the two constant dark energy equations of state have
also to agree. These analysis consider only
the effect of the growth factor and no other modification is considered. 
Another implementation performs a model-independent
reconstruction of the growth rate from distance measurements and then
compares to growth measurements~\cite{alam,mortonson}. Ref.~\cite{abate}
proposed a similar consistency test of the $\Lambda$CDM using low and high redshift SNIa
survey by estimating $\Omega_\mat$ in three different ways
(background geometry, growth, and shape of the power spectrum), all
agreeing with the canonical value 0.25. 

The original idea of Ref.~\cite{ub01} was extended to multiple
cosmological probes. Ref.~\cite{zhang07} proposed to use
the galaxy-velocity correlation and the galaxy-galaxy lensing,
which give access to $\langle\delta_g\theta_\mat\rangle\propto
bf\langle\delta_\mat^2\rangle$ and $\langle\delta_g\kappa\rangle\propto
b\langle\delta_\mat\Delta(\Phi+\Psi)\rangle$ so that the ratio
of these two quantities is expected to be independent of the biais,
at least in the regime of linear biasing. An estimator, $\hat E_G$
based on the ratio of these two quantities, was constructed and it
was demonstrated that it can distinguish a large class of models.
Ref.~\cite{jain07} worked out the relations between the various
observables, including a discussion of quasilinear effects. It
was also shown that the clustering of dark energy can mimic features of
a modification of GR and investigated the way to combine data
in order to distinguish the two effetcs. Refs.~\cite{songdore08,song08a} designed
consistency checks based on the redshift-space power spectrum
and weak lensing in order to constraint the ratio $\Phi/\Psi$
and the Poisson equation. Ref.~\cite{zhang08} proposed an estimator
to measure the ratio of the two gravitional potentials, again using
weak lensing and redshift-space power spectrum. Ref.~\cite{schmidt08}
proposed a method to extract the effect of a modified Poisson equation
and Ref.~\cite{guzik09} analyzed the combination of
imaging and spectroscopic surveys. Ref.~\cite{zhang06} proposed
to use the ISW-structure correlation to constrain the growth rate
of the density, and in particular its scale-dependence.

All these works are thus starting from the constitutive relations that exist
in a $\Lambda$CDM, and thus assuming GR valid to construct from
large scale structure survey some tests that will indicate the violation
of one of these relations. They often construct estimators that
are probed by using some extensions of GR (DGP, $f(R)$, scalar-tensor)
and in oder to forecast the power of coming surveys to distinguish between them.

\section{Conclusions}\label{sec:5}

This review has presented the tests of GR on astrophysical scales, but
more generally of the $\Lambda$CDM model, based on the large
scale structure and the global dynamics of the universe. In particular,
it is important to make the distinction since some supposedly tests
of GR proposed in the literature are in fact only tests of the 
$\Lambda$CDM model. Maybe the first answer these tests will give is
whether there is a need for new physical degrees of freedom in
our model and then start to caracterize the nature and the couplings
of this field with standard matter (and also dark matter).

As explained, future surveys will allow to map weak lensing, galaxy
distribution and velocity on sub-Hubble scales with high accuracy. This
will allow to construct many consistency checks of our cosmological model,
and in particular of GR. A multi-probe approach will allow
to have a better control of systematics which affect each probe.

Today, data shows no deviation from the $\Lambda$CDM model,
and thus from GR on large scales. On galactic
scales, the debate between dark matter and MOND-inspired models
is yet unsettled even though the need of massive neutrinos to reconcile
MOND with cluster data seem to disfavor this latter approach.
Besides, all analysis on cosmological still assume the existence
of dark matter.

It is important to keep in mind that these are not the only tests
of the deviation from the standard $\Lambda$CDM that can be performed.
Let us mention
\begin{itemize}
 \item{\it Test of the weak equivalence principle}. They can be performed
 on a large band of redshifts, up to BBN time, by constraining the time
 variation of fundamental constants~\cite{uzan03}.
 \item{\it Test of the distance duality relation}. In standard cosmology the
 angular and luminosity distances are related by $D_L=(1+z)^2D_A(z)$.
 This equation holds in any metric theory of gravity if the number
 of photons is conserved. By testing it~\cite{uzanam}, one can check the validity
 of Maxwell theory and constrain models such a photon-axion oscillation.
 \item{\it Test of the Copernican principle}. All the equations and solutions we have used,
 assumed the existence of a homogeneous and isotropic background spacetime.
 It is only an assumption based on the Copernican principle and recently many
 proposals to test it appear in the literature~\cite{uzanPC,stebbinsPC,chrisPC} .
 \item{\it Propagation of gravity waves}. In bi-metric theories of gravity, gravity
 waves and photons may  not necessary follow the geodesics of the same metric
 so that there can exist a time delay between them~\cite{kahya}. Confronting their
 arrival times (as well as those of neutrinos, if massive) allows to
 set constraints on bimetric theories of gravity. If gravity propagates 
 slower than light  then some tight constraint can arise from the
 energy loss of cosmic rays by gravitational \v{C}erenkov radiation~\cite{moore}
 leading to $c_{\rm GW}/c-1<2\times10^{-19}$.
\end{itemize}
These tests will enable to check the robustness of
the hypothesis on which our cosmological model rests. It will
either confirm the need for the existence of dark matter and dark energy
(thus extending drastically the domain of validity of GR)
or offer new theoretical constructions to explain the late time
acceleration of the cosmic expansion.

\begin{acknowledgements}
 It is a pleasure to thank my collaborators on this topic, 
 Francis Bernardeau, Gilles Esposito-Far\`ese, Yannick Mellier, Cyril Pitrou, Carlo Schimd.
\end{acknowledgements}


\end{document}